\documentclass[prb,twocolumn, superscriptaddress]{revtex4-1}

\usepackage{graphicx}
\usepackage{dcolumn}
\usepackage{bm}

\begin{document}


\title{Giant Rashba splitting of quasi-1D surface states of Bi/InAs(110)-(2$\times$1)}

\author{Takuto Nakamura}
\affiliation{Department of Physics, Graduate School of Science, Osaka University, Toyonaka 560-0043, Japan}
\author{Yoshiyuki Ohtsubo}
\email{y\_oh@fbs.osaka-u.ac.jp}
\affiliation{Graduate School of Frontier Biosciences, Osaka University, Suita 565-0871, Japan}
\affiliation{Department of Physics, Graduate School of Science, Osaka University, Toyonaka 560-0043, Japan}
\author{Yuki Yamashita}
\affiliation{Department of Physics, Graduate School of Science, Osaka University, Toyonaka 560-0043, Japan}
\author{Shin-ichiro Ideta}
\author{Kiyohisa Tanaka}
\affiliation{Institute for Molecular Science, Okazaki 444-8585, Japan}
\author{Koichiro Yaji}
\author{Ayumi Harasawa}
\author{Shik Shin}
\author{Fumio Komori}
\affiliation{Institute for Solid State Physics, The University of Tokyo, 5-1-5 Kashiwanoha, Kashiwa, Chiba 277-8581, Japan}
\author{Ryu Yukawa}
\author{Koji Horiba}
\author{Hiroshi Kumigashira}
\affiliation{Photon Factory, Institute of Materials Structure Science, High Energy Accelerator Research Organization (KEK), 1-1 Oho, Tsukuba 305-0801, Japan}
\author{Shin-ichi Kimura}
\email{kimura@fbs.osaka-u.ac.jp}
\affiliation{Graduate School of Frontier Biosciences, Osaka University, Suita 565-0871, Japan}
\affiliation{Department of Physics, Graduate School of Science, Osaka University, Toyonaka 560-0043, Japan}

\date{\today}

\begin{abstract}
Electronic states of the Bi/InAs(110)-(2$\times$1) surface and its spin-polarized structure are revealed by angle-resolved photoelectron spectroscopy (ARPES), spin-resolved ARPES, and density-functional-theory calculation.
The surface electronic state showed quasi-one-dimensional (Q1D) dispersion curves and a nearly metallic character; the top of the hole-like band is just below the Fermi level.
The size of the Rashba parameter ($\alpha_{\rm R}$) reached much larger value ($\sim$5.5 eV\AA) than previously reported 1D systems..
The present result would provide a fertile playground for further studies of the exotic electronic phenomena in 1D or Q1D systems with the spin-split electronic states as well as for advanced spintronic devices.
\end{abstract}

\maketitle

\section{Introduction}
Spin-polarized electronic states without ferromagnetic materials, such as the spin-split low-dimensional states due to the Rashba effect \cite{Rashba} and topological surface states \cite{Hasan10}, have been studied extensively for future application to spintronics \cite{Manchon15}.
Among them, one-dimensional (1D) and quasi-1D (Q1D) states are regarded as promising systems because of various merits such as possible downsizing of the system and highly efficient suppression of backscattering.
Recently, spin-split (Q)1D states have attracted much attention in basic science research to realize exotic electronic phenomena such as Majorana bound states \cite{Mourik12} and spin-dependent density-wave formation \cite{Braunecker10}.

\begin{figure}
\includegraphics[width=80mm]{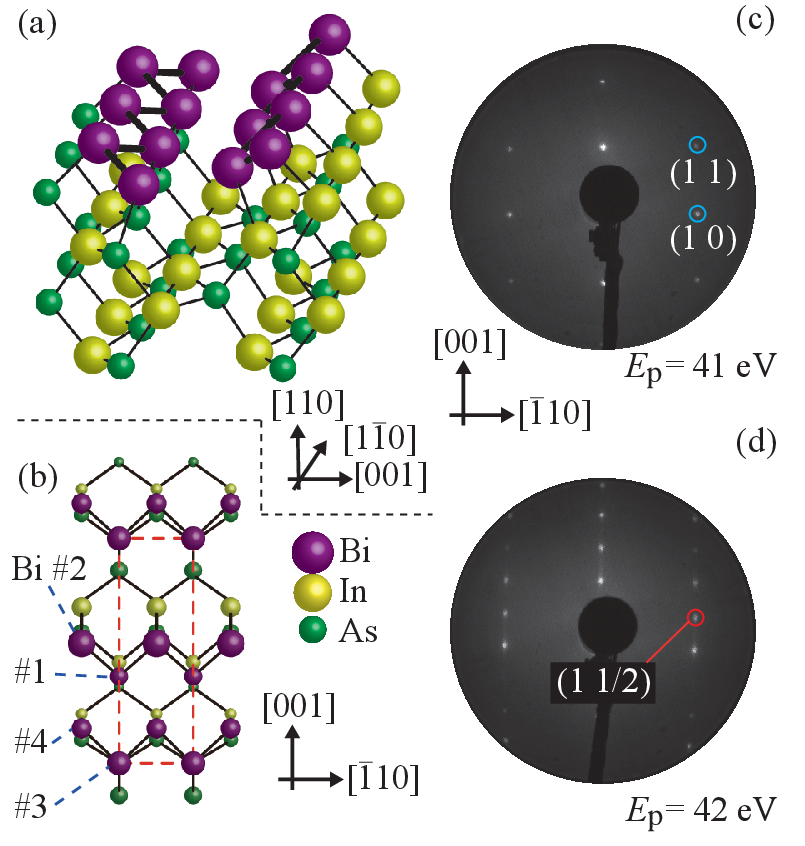}
\caption{\label{fig1} (Color online).
(a, b) Surface atomic structure of Bi/InAs(110)-(2$\times$1) \cite{Betti99}. The dashed rectangle in (b) is the (2$\times$1) surface unit cell.
(c, d) Low-energy electron diffraction patterns of (c) a cleaved InAs(110) surface and (d) the Bi/InAs(110)-(2$\times$1) surface.
Both patterns were taken at room temperature.
}
\end{figure}

Rashba-type spin-orbit interaction (SOI) causes spin-split 2D (or 1D) electronic states in a low-dimensional system without space-inversion symmetry, \textit{e.g.}, in crystal surfaces.
Since it causes no ferromagnetic order, time-reversal symmetry remains even with spin-split states.
Assuming a parabolic dispersion with spin splitting proportional to the wavevector $k$ from the Kramers degeneracy point, the size of typical Rashba spin splitting can be expressed as $\Delta E = 2\alpha_{\rm R}k$, where the proportionality constant $\alpha_{\rm R}$ is the so-called Rashba parameter, which is often used as a scale to estimate the size of Rashba SOI.
Both theoretical and experimental effort for the spintronic application of such Rashba-split states have been made from earlier days \cite{Datta90}.
In these days, giant Rashba spin splitting has been realized in both 2D \cite{Nakagawa07, Ast07, Miyamoto14, Chen15, Liebmann16} and 3D \cite{Ishizaka11} systems, where $\alpha_{\rm R}$ is 3-7 eV\AA\, and spin-to-current conversion has been realized as the first step toward practical spintronic application using such Rashba-split states \cite{Rojas13}.
However, despite the extensive studies and fascinating theoretical prospects introduced above, a quest for (Q)1D states with giant Rashba splitting has been the challenge for surface science in this dacade, after the discovery of the 2D ones \cite{Nakagawa07, Ast07}.
Although some Q1D systems with Rashba splitting are known, such as Au and Pb on vicinal Si surfaces \cite{Barke06, Okuda10, Yeom14, Kopciuszynski17}, Pt/Si(110) \cite{Park13}, the 1D step edge of Bi thin films \cite{Takayama15}, and Pt/Ge(001) \cite{Yaji16_Pt}, their Rashba parameters are at most $\sim$1.4 eV\AA, which is much smaller than those of the giant-Rashba counterparts in 2D and 3D systems.

Various theoretical studies to achieve a larger Rashba SOI have been reported so far \cite{Bihlmayer06, Premper07, Oguchi09, Nagano09, Park11, Bahramy11, Krasovskii14, Ishida14}, and many parameters have been proposed as the origin of the giant Rashba effect, such as the asymmetric charge distribution in the proximity of surface atoms \cite{Bihlmayer06, Nagano09}, orbital-angular-momentum polarization \cite{Park11}, and hybridization between different atomic orbitals \cite{Premper07, Krasovskii14, Ishida14}.
Although it is difficult to reach a firm conclusion from such a long debate, some clues can be obtained from these studies.
Firstly, the target electronic state should be derived from heavy elements whose nucleus potential causes a large SOI.
Secondly, the state should be quite different from nearly-free-electron (NFE) isotropic wavefunctions, since the Rashba SOIs for such NFE states are simulated to be very small by using all the theoretical models.

The Bi/InAs(110)-(2$\times$1) surface has been extensively studied in the 1990s as a typical ordered metal-semiconductor interface \cite{Betti98, Betti99, Renzi99}.
The detailed surface atomic structure has been determined by surface X-ray diffraction \cite{Betti99} to be strongly buckled zig-zag Bi chains, as shown in Fig. 1 (a).
Previous studies also reported a small density of states around the Fermi level ($E_{\rm F}$), suggesting metallic surface states \cite{Renzi99}.
This surface is formed by heavy Bi atoms and has a Q1D atomic structure, implying anisotropic surface states away from NFE case.
Therefore, Bi/InAs(110)-(2$\times$1) is one of the promising surfaces to realize large Rashba-type spin splitting in a Q1D system.
However, despite these properties, no research has been reported so far about the detailed surface-band dispersion, especially focused on possible spin-split structures, of the Bi/InAs(110)-(2$\times$1) surface.

In this paper, we report on the surface states of Bi/InAs(110)-(2$\times$1) and its spin-polarized structure through angle-resolved photoelectron spectroscopy (ARPES), spin-resolved ARPES (SARPES), and density-functional-theory (DFT) calculation.
The surface state showed highly anisotropic, Q1D dispersion around $E_{\rm F}$ and a nearly metallic (semiconducting) character; the top of the hole-like surface band is just below $E_{\rm F}$ ($\sim$50 meV).
The size of $\alpha_{\rm R}$ reached quite a large value ($\sim$5.5 eV\AA).
The results provide a promising foundation for studying the spin-split Q1D electronic states as well as for developing advanced spintronic devices.

\section{Methods}
A clean InAs(110) surface was prepared by cleaving the side face of InAs(001) substrates (nominally undoped) in ultra-high-vacuum chambers.
As shown by the sharp and low-background low-energy electron diffraction (LEED) pattern in Fig. 1 (c), clean and well-defined (110) surfaces were obtained by cleavage.
Then, Bi were evaporated from a Knudsen cell at room temperature.
Subsequent annealing at 563 K for 15 min produced the (2$\times$1) surface, as indicated by the LEED pattern in Fig. 1 (d).
The sample preparation procedure is the same as that reported in previous studies \cite{Betti98, Renzi99}.

ARPES measurements were performed at BL-2A MUSASHI of the Photon Factory and BL5U of UVSOR-III with photon energies ranging from 50 to 85 eV.
SARPES measurements were performed at the Institute for Solid State Physics, the University of Tokyo with linearly polarized photons by using a laser source ($h\nu$ = 6.994 eV) \cite{Yaji16}.
Spin-integrated ARPES data were taken at the same time with both linearly and circularly polarized photons by laser for the sake of comparison.
For the SARPES measurements, the photon-incident plane was (001), and the electric-field vector of the photons was normal to the incident plane (001).
The effective Sherman function of the spin detector was 0.27.
The energy resolutions and the positions of $E_{\rm F}$ of both spin-integrated and spin-resolved ARPES were calibrated by the Fermi edge of a Cu block touching the sample.
The energy resolutions of both measurements were evaluated to be $\sim$20 meV. 

DFT calculation was performed using the WIEN2k code with SOI taken into account\cite{W2k}.
The Engel-Vosko generalized gradient approximation \cite{Engel93} was utilized to construct the exchange and correlation potentials.
The surface was modeled by an asymmetric slab of 10 InAs layers with the surface covered by (2$\times$1) zig-zag Bi chains.
The surface atomic structure was energetically optimized down to the third In and As layers; the bottom of the slab was truncated from the bulk InAs structure and terminated by hydrogen atoms.

\begin{figure*}
\includegraphics[width=150mm]{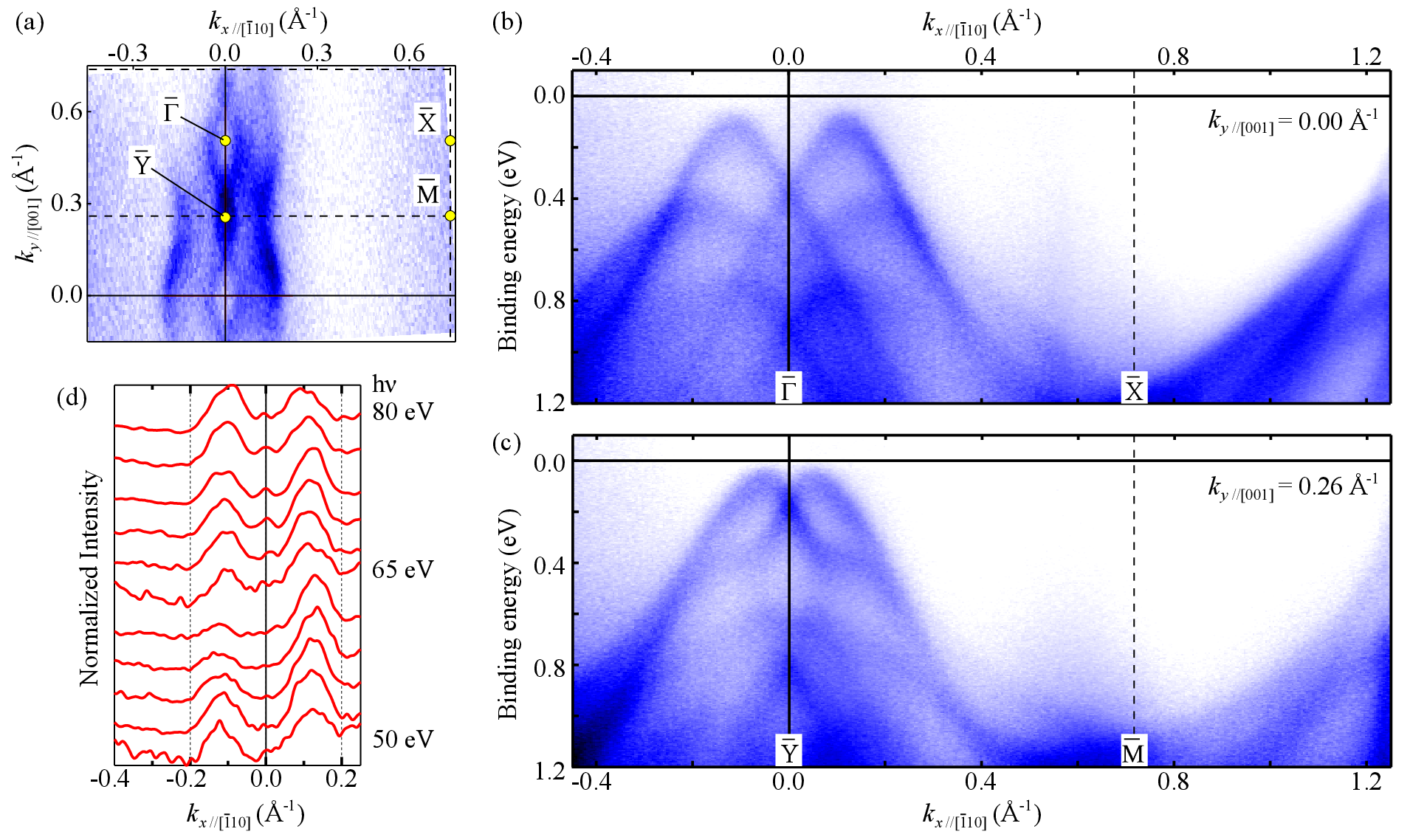}
\caption{\label{fig2} (Color online).
Electronic structure of Bi/InAs(110)-(2$\times$1) measured by ARPES at 15 K with circularly polarized photons. (a) A constant energy contour at a binding energy of 200  $\pm$ 10 meV ($h\nu$ = 55 eV). The dashed lines indicate boundaries of the (2$\times$1) surface Brillouin zone. 
(b, c) ARPES intensity plots along (b) $\bar{\Gamma}$-$\bar{\rm X}$ ($k_y$ = 0.00 \AA$^{-1}$) and (c) $\bar{\rm Y}$-$\bar{\rm M}$ ($k_y$ = 0.26 \AA$^{-1}$) taken with 55 eV photons. 
(d) Momentum-distribution curves at a binding energy of 100 $\pm$ 20 meV taken with photon energies ranging from 50 to 80 eV at $k_y$ = 0.00 \AA$^{-1}$.
}
\end{figure*}

\section{Results and Discussion}

\subsection{Surface electronic structure of Bi/InAs(110)-(2$\times$1)}
Figure 2 (a) shows the ARPES constant-energy contour at the binding energy of $200\pm 10$ meV.
$k_x$ ($k_y$) is defined parallel to [$\bar{1}$10] ([001]).
The contour shows an anisotropic, Q1D shape with an open, waving contour along [001].
A waving feature in the energy contour indicates a small but finite 2D interaction between surface atomic chains.
Figures 2 (b) and 2 (c) show ARPES intensity plots along $\bar{\Gamma}$-$\bar{\rm X}$ and $\bar{\rm Y}$-$\bar{\rm M}$, respectively.
The top of the hole bands lies slightly below $E_{\rm F}$, indicating a semiconducting character. 
In general, such surface states lying around $E_{\rm F}$ agree well with previous studies \cite{Betti98, Renzi99}.

The only difference is that these surface electronic states were reported to be metallic in the previous studies.
However, the same (2$\times$1) LEED patterns and similar spectral shapes (Fig. 3 (c)) to those reported in refs. \cite{Betti98, Renzi99} suggest that the surface atomic structures are the same.
One possible reason for the discrepancy in the surface electronic states is that all the ARPES and SARPES measurements in this work were performed at low temperatures (15 to 40 K), in contrast to those at room temperature in the previous studies.
At room temperature, the surface accumulation layer discussed above might supply carriers (holes in this case) to the surface states, making them metallic.
These results are quite encouraging for future works, especially on spin-dependent transport phenomena via the spin-split surface states on Bi/InAs(110)-(2$\times$1), because the difference in results between the previous works and this work indicate that the carrier amount in the surface state could be tuned by some properties such as sample temperature.

Figure 2 (d) displays momentum distribution curves (MDCs) at a binding energy of 100 $\pm$ 20 meV taken in a photon energy range from 50 to 80 eV at  $k_{y}$ = 0.00 \AA$^{-1}$. 
The two peaks at  $k_{x}$ =  $\pm$0.11\AA$^{-1}$ correspond to the hole bands around $\bar{\Gamma}$ in Fig. 2 (b).
The peak positions of hole bands do not change depending on the photon energies, indicating that the Q1D bands are surface states without any dispersion along the surface normal.
In addition, a weak intensity appears around $k_x$ = 0.00 \AA$^{-1}$ at $h\nu >$ 65 eV.
It would be due to the bottom of bulk conduction band with 3D dispersion, as shown below.

\begin{figure}[htbp]
\includegraphics[width=80mm]{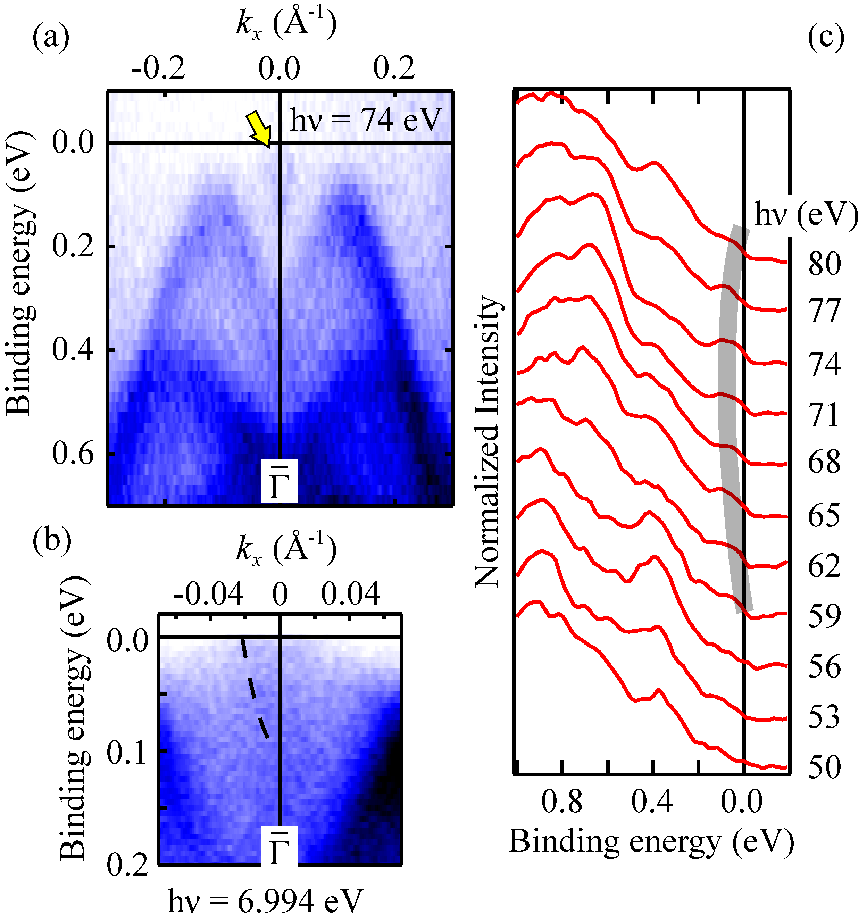}
\caption{\label{figs3}
(a) ARPES intensity plot along $\bar{\Gamma}$-$\bar{\rm X}$ taken at $h\nu$ = 74 eV at 15 K.
(b) A close-up image of the ARPES intensity plots around $\bar{\Gamma}$ taken by laser ($h\nu$ = 6.449 eV) at 40 K.
(c) ARPES energy distribution curves around $\bar{\Gamma}$ (integrated within $|k_x| < $ 0.025 \AA$^{-1}$) with various photon energies.
A fat curve is a guide to the eye of the small electron pocket.
}
\end{figure}

Figure 3 (a) shows ARPES intensity plots around $\bar{\Gamma}$ taken with 74 eV.
As indicated by the arrow, a metallic band forming a small electron pocket at $\bar{\Gamma}$ is observed with $h\nu$ = 74 eV but is absent with $h\nu$ = 55 eV (see Fig. 2 (b)).
The small momentum-distribution-curve (MDC) peaks at $k_x$ = 0 \AA$^{-1}$ in Fig. 2 (d) ($h\nu > $ 65 eV) correspond to the bottom of this small electron pocket.
In addition, a similar electron-pocket-like feature was observed in the ARPES intensity maps taken with a laser source (h$\nu$ = 6.994 eV, Fig. 3 (b)).
In order to trace this feature, the energy distribution curves at $\bar{\Gamma}$ with photon energies from 50 to 80 eV are shown in Fig. 3 (c).
This metallic feature appears at $h\nu$ = 59 eV at $E_{\rm F}$ and shows small dispersion depending on the photon energies, as indicated by a fat line in Fig. 3 (c).
As shown there, this feature has the bottom at $h\nu \sim$ 70 eV (around 0.1 eV) and lies nearly at $E_{\rm F}$ again at $h\nu$ = 80 eV.
Note that this feature is observed in wider energy range than those in Fig. 2 (d) because the momentum distribution curves in Fig. 2 (d) is taken at 100 meV below $E_{\rm F}$.
This clear dependence on the incident photon energy indicates that this small electron pocket is a 3D band; perhaps a bottom of the 3D, bulk conduction band.
Although the InAs substrates used in this work are nominally undoped, the surface preparation procedure, Bi evaporation, and subsequent annealing might form a $p$-type surface accumulation layer.

\subsection{Spin splitting of the surface states}

\begin{figure}[htbp]
\includegraphics[width=80mm]{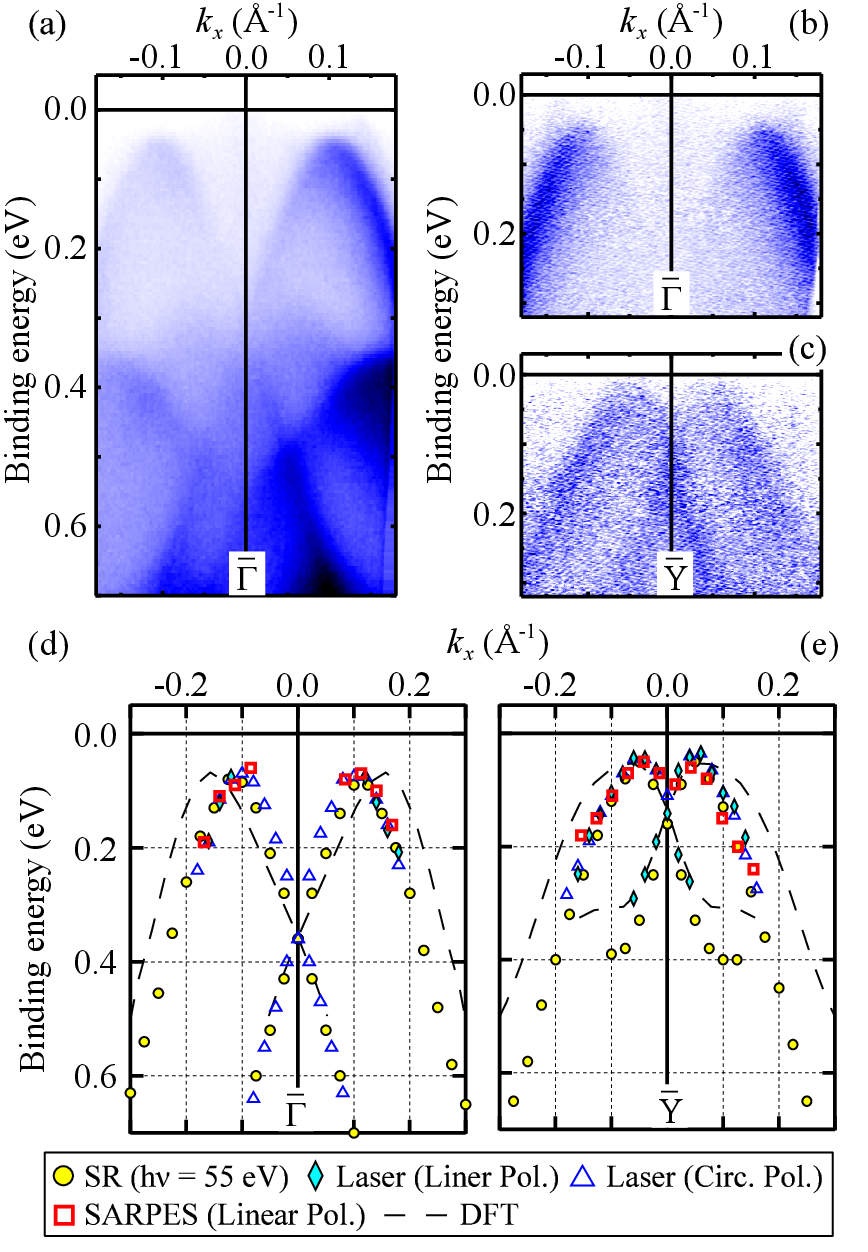}
\caption{\label{figs2} 
(a-c) ARPES intensity plots at 40 K by laser ($h\nu$ = 6.449 eV) with (a) circularly polarized photons along $\bar{\Gamma}$-$\bar{\rm X}$, (b) linearly polarized photons along $\bar{\Gamma}$-$\bar{\rm X}$, and (c) linearly polarized photons along $\bar{\rm Y}$-$\bar{\rm M}$.
(d) The overlap of the surface band dispersions obtained from ARPES (with both synchrotron radiation (SR) and laser), SARPES and DFT along $\bar{\Gamma}$-$\bar{\rm X}$. The surface band from DFT is energetically shifted so that the top of the parabola lies at the same energy as those from ARPES.
(e) The same as (d) but along $\bar{\rm Y}$-$\bar{\rm M}$.
}
\end{figure}

A pair of parabolic surface bands separating with respect to $k_x$ = 0 \AA$^{-1}$ suggests the spin splitting due to Rashba-type SOI \cite{Rashba}.
In order to elucidate the spin texture of the surface states, the SARPES measurements is desirable.
Before making the spin-polarization analysis, the correspondence of the spin-integrated and spin-resolved ARPES data with each other.
Figures 4 (a-c) show spin-integrated ARPES intensity plots by using a laser source ($h\nu$ = 6.994 eV).
The obtained dispersion of the paired parabolic bands below $E_{\rm F}$ agrees well with those observed by higher photon energies.
With linearly polarized photons, the photoelectron intensity in the vicinity of $\bar{\Gamma}$ decreases.
It would be owing to the photoexcitation selection rule.
In Figs. 4 (d, e), the ARPES peak positions obtained by synchrotron radiation ($h\nu$ = 55 eV), laser (both spin-resolved and spin-integrated ones) and DFT calculations (from Fig. 6 in the next section) are plotted together.
All the ARPES and SARPES peaks agree well, suggesting that they originate from the same surface bands.

\begin{figure}
\includegraphics[width=70mm]{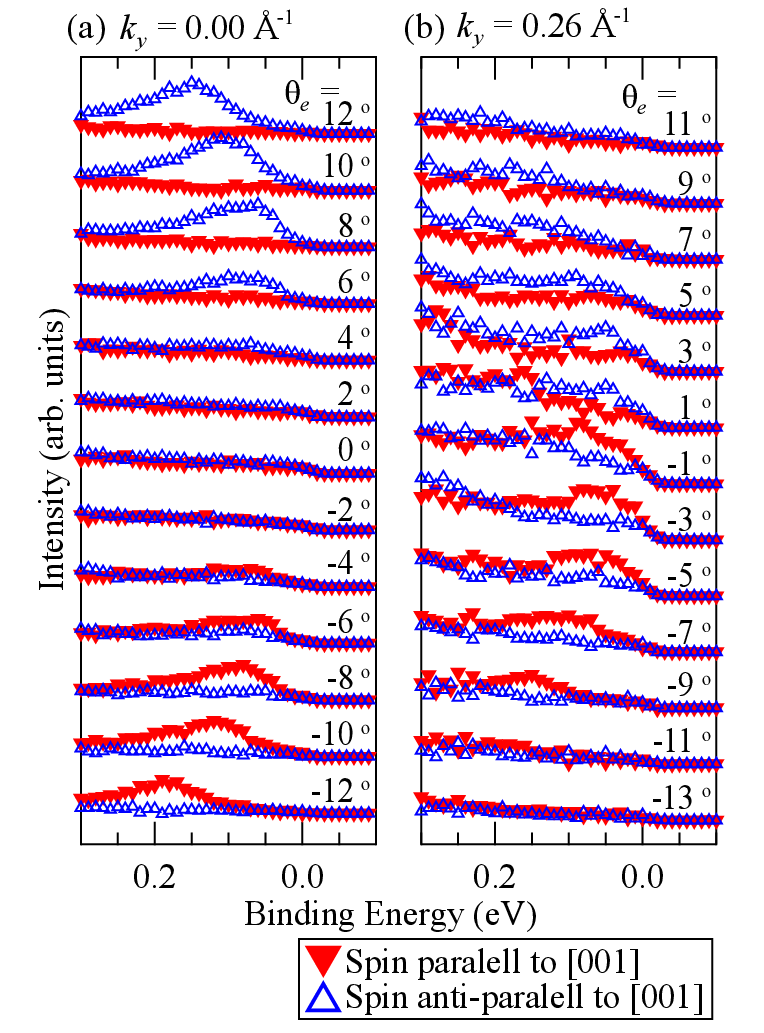}
\caption{\label{fig3} (Color online).
SARPES energy distribution curves taken along [1$\bar{1}$0] ($k_x$) at (a) ($k_{y//[001]}$ = 0.00 \AA$^{-1}$  and (b)  ($k_{y//[001]}$ = 0.26 \AA$^{-1}$) at 40 K. $\theta_e$  = 10 $^\circ$  corresponds to 0.14  \AA$^{-1}$.
}
\end{figure}

Figure 5 shows the SARPES energy distribution curves taken along  $\bar{\Gamma}$-$\bar{\rm X}$ (a) and  $\bar{\rm Y}$-$\bar{\rm M}$ (b) at 40 K.
The filled (open) triangles correspond to spin polarization parallel (anti-parallel) to [001], the in-plane orientation perpendicular to $k_x$.
The spin-polarized peaks in Figs. 3 (a, b) disperse downwards from $k_x$ = 0 \AA$^{-1}$.
The spin-polarization orientations invert together with the sign of the emission angles (nearly proportional to $k_x$), and the signs of the polarizations are the same along both $\bar{\Gamma}$-$\bar{\rm X}$ and $\bar{\rm Y}$-$\bar{\rm M}$.

\subsection{Surface states obtained by DFT calculation}

Figure 6 shows the calculated surface electronic structure of the Bi/InAs(110)-(2$\times$1) surface.
A parabolic hole-like dispersion, the top of which is around $E_{\rm F}$, is obtained, showing good qualitative agreement with the states observed by ARPES, even though the experimental $E_{\rm F}$ is pinned near the bottom of the conduction band at $\bar{\Gamma}$.
Although the quantitative agreement is not perfect, as shown in Figs. 4 (d) and 4 (e), such small difference often occurs between DFT and ARPES due to various inperfection of DFT calculation, such as thin slab thickness and a band-gap misestimation by exchange-correlation functional.

The spin-polarized character of the calculated surface states also agrees well with the SARPES results.
The spin polarization of the surface states decreases significantly in the vicinity of $\bar{\Gamma}$ and $\bar{\rm Y}$.
This could be due to the influence of other underlying surface states ($\sim$0.4 eV at $\bar{\Gamma}$ and $\bar{\rm Y}$), which also have a large contribution from the surface Bi atoms with spin orientations opposite to those of the ``upper'' states.
Such spin-orbital interference as well as the decrease of the spin polarization of the ``upper'' branch of the Rashba-split bands have been already reported in the 2D giant-Rashba systems Bi/Ag(111) and Pb/Ag(111) and discussed in detail \cite{Bihlmayer07}.

\begin{figure}
\includegraphics[width=80mm]{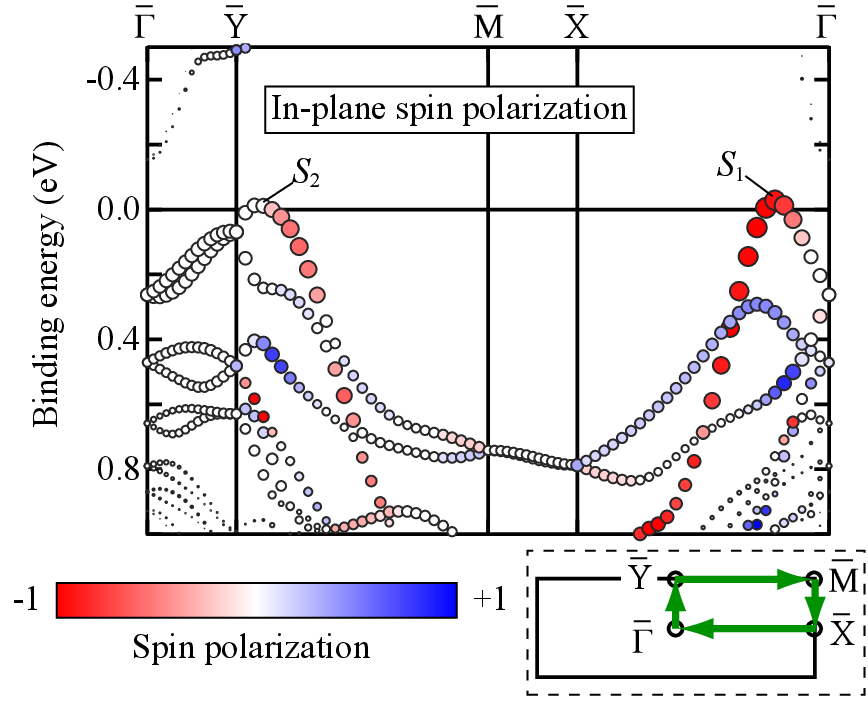}
\caption{\label{fig4} (Color online).
Calculated band structure in the (2$\times$1) surface Brillouin zone. The wave-vector path is indicated in the inset. The band structure was calculated for 10 InAs layers covered with Bi zig-zag chains.
The radii of the circles are proportional to the contribution from the atomic orbitals of the surface Bi.
The contrasts (colors) of each circle represent the spin polarizations along in-plane directions, with the negative values (red) corresponding to the spins parallel to [001].
}
\end{figure}

\subsection{Giant Rashba effect on Bi/InAs(110)}
The paired parabolic dispersion as well as the in-plane spin polarizations without breaking time-reversal symmetry, which are perpendicular to both the surface normal and the dispersion direction $k_x$, strongly suggest Rashba-type spin splitting.
The size of the Rashba parameter $\alpha_{\rm R}$ for parabolic bands can be estimated as $\alpha_{\rm R}$ = 2$E_{\rm R}$/$k_0$ \cite{Ast07}, where $E_{\rm R}$ and $k_0$ are the energy and wavenumber differences, respectively, between the top of the hole-like band and the Kramers degenerate point ($\bar{\Gamma}$ or $\bar{\rm Y}$ in the current case, see also Fig. 7).
As shown in Figs. 7 (a) and 7 (b), the surface bands on Bi/InAs(110)-(2$\times$1) can be fit well with a parabola.
From the surface-band dispersion (Figs. 7 (a, b) and Fig. 2), the Kramers degeneracy points (crossing points) along $k_x$ are 0.36 (0.14) eV at $\bar{\Gamma}$ ($\bar{\rm Y}$) and the top of the surface bands lies at 0.07 (0.04) eV at $k_x$ = $\pm$0.105 (0.055) \AA$^{-1}$ around $\bar{\Gamma}$ ($\bar{\rm Y}$).
Note that these $k_x$ values can be used as $k_0$ in the equation above.
From these values obtained by ARPES, $E_{\rm R}$ along $\bar{\Gamma}$-$\bar{\rm X}$ is 0.29 eV, resulting in $\alpha_{\rm R}$ = 5.5 eV \AA.
By the same calculation, ($E_{\rm R}$, $k_0$) = (0.10 eV, 0.055 \AA$^{-1}$) along $\bar{\rm Y}$-$\bar{M}$; thus, $\alpha_{\rm R}$ = 3.6 eV \AA.
To our knowledge, the maximum value of $\alpha_{\rm R}$ in this system (5.5 eV \AA) is 4-5 times larger than those of other 1D or Q1D Rashba systems \cite{Park13, Takayama15}.
Even when including 2D and 3D systems, the maximum value is greater than those of the typical $giant$ Rashba systems such as Bi/Ag(111) \cite{Ast07}, GeTe(111) \cite{Liebmann16} and BiTeI \cite{Ishizaka11} and is the largest among the Rashba-split states lying around $E_{\rm F}$, which is an important character to realize spin-dependent transport phenomena.
Only few surface states far below $E_{\rm F}$ \cite{Miyamoto14, Chen15} have larger $\alpha_{\rm R}$ values than that in the current case.
%
\begin{figure}
\includegraphics[width=80mm]{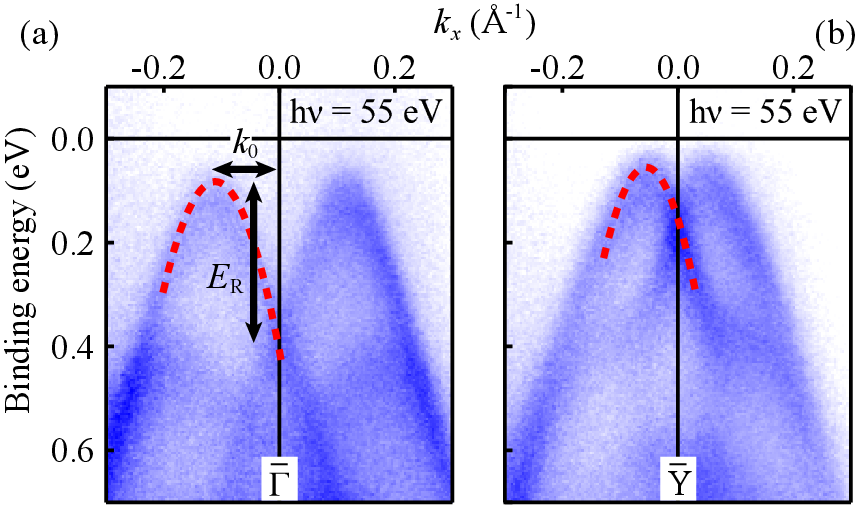}
\caption{\label{figs1}
ARPES intensity plots from Bi/InAs(110)-(2$\times$1) at 15 K along (a) $\bar{\Gamma}$-$\bar{\rm X}$ and (b) $\bar{\rm Y}$-$\bar{\rm M}$. The dashed lines are the guides of the parabolic dispersion of the hole bands. The arrows in (a) show the definition of $E_{\rm R}$ and $k_0$ for estimation of the Rashba parameter (see text for details). 
}
\end{figure}

\begin{table}
\caption{
Calculated fractional contributions of atomic orbitals of surface Bi to the spin-split surface states ($S_1$ and $S_2$ in Fig. 6) and the geometric means between 6$p$ orbitals M$_{xy}$, M$_{yz}$ and M$_{zx}$).
Atom numbers are indicated in Fig. 1 (b).
Each value is normalized by 6$p_y$ of $S_1$ (Bi \#1). 
The $p_x$, $p_y$, and $p_z$ orbitals correspond to the $p$ orbitals, the lobes of which are along [$\bar{1}$10], [001], and [110], respectively.
}
\label{tab1}
\begin{ruledtabular}
\begin{tabular}{cccccccc}
$S_1$ ($\bar{\Gamma}$-$\bar{\rm X}$) & 6$s$ & 6$p_x$ & 6$p_y$ & 6$p_z$ & M$_{xy}$ & M$_{yz}$ &M$_{zx}$\\
\hline
\#1 &0.20 &0.20 &1 &0.04 & 0.45 & 0.19 & 0.09 \\
\#2 &0.02 &0.67 &0.20 &0.54 & 0.36 & 0.33 & 0.60 \\
\#3 &0.01 &0.62 &0.11 &0.47 & 0.27 & 0.23 & 0.54 \\
\#4 &0.27 &0.23 &0.76 &0.02 & 0.42 & 0.12 & 0.07 \\
sum & & & & & 1.50 & 0.87 & 1.30 \\
\hline
$S_2$ ($\bar{\rm Y}$-$\bar{\rm M}$) & 6$s$ & 6$p_x$ & 6$p_y$ & 6$p_z$ & M$_{xy}$ & M$_{yz}$ &M$_{zx}$\\
\hline
\#1 &0.18 &0.64 &0.65 &0.00 & 0.64 & 0.00 & 0.00 \\
\#2 &0.00 &0.83 &0.41 &0.25 & 0.58 & 0.32 & 0.46 \\
\#3 &0.00 &0.32 &0.24 &0.53 & 0.28 & 0.36 & 0.41 \\
\#4 &0.27 &0.31 &0.29 &0.01 & 0.30 & 0.04 & 0.04 \\
sum & & & & & 1.80 & 0.72 & 0.91 \\
\hline
\end{tabular}
\end{ruledtabular}
\end{table}

The fractional contributions of Bi atomic orbitals to the spin-split surface states are listed in Table I.
The main contribution to these states are from the 6$s$ and 6$p$ orbitals of the surface Bi atoms.
Owing to the major contribution from 6$p$, the eigenfunctions of the surface states are quite different from NFE ones, justifying the giant size of the Rashba effect.

The maximum of the Rashba parameter $\alpha_{\rm R}$ is along $\bar{\Gamma}$-$\bar{\rm X}$ (5.5 eV \AA).
Although it is still ``giant'' along $\bar{\rm Y}$-$\bar{\rm M}$, the value (3.6 eV \AA) is smaller than that along $\bar{\Gamma}$-$\bar{\rm X}$.
This could be due to the slight modifications of the surface Bi orbitals.
According to a recent theoretical study \cite{Ishida14}, the size of the spin splitting is proportional to the geometric means between atomic orbitals having the same orbital characters but different magnetic quantum numbers: $p_x$, $p_y$, and $p_z$ in this case.
To trace this feature, M$_{ab}$, a geometric mean of fractional contributions between Bi 6$p_a$ and 6$p_b$ ($a, b = x, y, z$), is calculated as shown in Table I.
Since the spin-split states of Bi/InAs(110)-(2$\times$1) have sizable fractional contributions from all 6$p$ orbitals, all the geometric means (M$_{xy}$, M$_{yz}$ and M$_{zx}$ in Tab. I) have finite values.
From their sum (see Table I), both ${\rm M}_{yz}$ and ${\rm M}_{zx}$ are suggested to be the scale of the spin splitting; ${\rm M}_{yz}(S_1)$/${\rm M}_{yz}(S_2)\sim$ 1.2, ${\rm M}_{zx}(S_1)$/${\rm M}_{zx}(S_2)\sim$ 1.4, and $\alpha_{\rm R}(S_1)$/$\alpha_{\rm R}(S_2)\sim$ 1.5.
These are the geometric means between in-plane and out-of-plane 6$p$ orbitals.
Such hybridizations between in-plane and out-of-plane orbitals have also been observed in other giant Rashba systems \cite{Bihlmayer07, Miyamoto14, Chen15, Premper07, Nagano09, Bahramy11, Ishida14}.
In addition, it should be noted that the nonequivalence between $\bar{\Gamma}$ and $\bar{\rm Y}$ originates from the 2D, inter-Bi-chain interaction, which suggests that external-field application normal to the 1D chain could tune the size of the 1D Rashba effect.

The giant spin splitting in the Q1D surface state discovered here is expected to be a promising template for future spintronic devices \cite{Manchon15}.
For example, the efficiency of spin-to-charge conversion by the inverse Edelstein effect (IEE), which is scaled by $\lambda _{IEE}$, is estimated to be $\sim$ 4 nm for this state, assuming the momentum relaxation time of 5 fs: this relaxation time is nearly an average of known surface-Rsahba systems \cite{Rojas13, Karube16, Han18}.
This $\lambda _{IEE}$ value is an order of magnitude larger than those of 2D Rashba systems \cite{Rojas13, Karube16} and even larger than those of topological insulators \cite{Han18}, implying that sizable spin-dependent transport in 1D should be observed on Bi/InAs(110).
In addition, for such applications, a method to control surface carriers is desirable.
Such control would be achieved easily for this surface state because some previous studies actually reported metallic surface states \cite{Renzi99}.
For tuning the surface carrier, carrier accumulation depending on the substrate temperature as discussed in Section II as well as surface alloying with smaller-valence elements such as Pb would be applicable.
In addition to industrial applications, the fine tuning of the surface Q1D state with spin splitting is expected to cause exotic electronic phenomena that would be of interest in basic science studies, such as studies on spin-dependent density-wave formation \cite{Braunecker10} and to realize Majorana bound states \cite{Mourik12}.

\section{Summary and Conclusion}
In summary, we have investigated the quasi-one-dimensional (Q1D) surface states on the Bi/InAs(110)-(2$\times$1) surface and its spin-polarized structure by angle-resolved photoelectron spectroscopy (ARPES), spin-resolved ARPES, and density-functional-theory (DFT) calculation.
The surface state showed a nearly metallic character, lying just below the Fermi level ($E_{\rm F}$) with giant Rashba spin splitting whose the Rashba parameter $\alpha_{\rm R}$ reached $\sim$5.5 eV\AA.
This value of the Rashba parameter is 4-5 times larger than that of other 1D or Q1D Rashba systems and is the largest value among all the experimentally realized Rashba-split systems around $E_{\rm F}$.
The present results are expected to provide a promising foundation for further studies of the exotic electronic phenomena in 1D or Q1D systems with spin-split electronic states as well as for the development of advanced spintronic devices.

\section*{Ackowledgements}
We thank J.-C. Rojas-S$\acute{\rm a}$nchez for the helpful discussions.
We also thank Y. Takeno for his support during general experiments.
The ARPES measurements were partially performed under UVSOR proposal 29-534 and Photon Factory proposal 2017G537.
The SARPES experiment in this work was carried out by joint research in ISSP, the University of Tokyo.
This work was also supported by JSPS KAKENHI (Grants Nos. JP26887024 and JP17K18757).


\begin{figure}
\includegraphics[width=80mm]{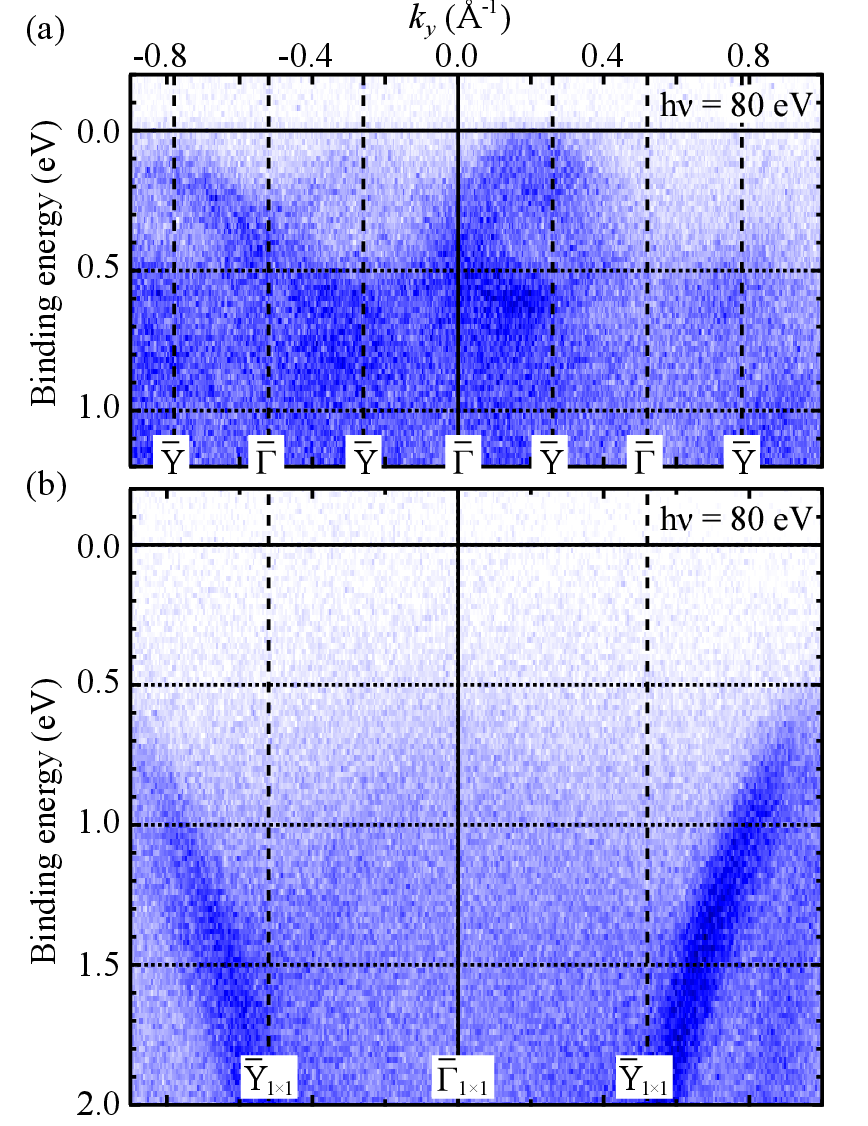}
\caption{\label{figs4}
ARPES intensity plots at 15 K ($h\nu$ = 80 eV) with linearly polarized photons, taken from (a) Bi/InAs(110)-(2$\times$1) and (b) as-cleaved pristine InAs(110) surface. Energy resolution of these plots were 40 meV.
}
\end{figure}

\subsection*{Appendix A: Surface band dispersion and Rashba spin splitting normal to the zig-zag Bi chain}
Figure 8 (a) shows an ARPES intensity plot along $\bar{\Gamma}$-$\bar{\rm Y}$, the surface-band dispersion perpendicular to the surface zig-zag Bi chain.
It shows a clear periodicity according to the (2$\times$1) surface Brillouin zone.
This dispersion between $\bar{\Gamma}$ and $\bar{\rm Y}$ is derived from finite 2D interaction across the surface Bi chain.
The overall feature of this dispersion agrees well with the DFT calculation shown in Fig. 6.
The Rashba splitting of the surface band is not resolved along this direction, because the splitting is smaller than the peak width of the ARPES energy distribution curves ($\sim$100 meV).
The small size of the Rashba splitting might have resulted from the strong suppression of splitting due to the surface mirror plane along $\bar{\Gamma}$-$\bar{\rm Y}$ \cite{Ohtsubo12}.

\subsection*{Appendix B: Comparison with a pristine InAs(110) surface}
To reveal the role of the (2$\times$1)-Bi surface atomic structure to the surface electronic states, the band dispersison along $\bar{\Gamma}$-$\bar{\rm Y}$ from the as-cleaved, pristine (110) surface of InAs is shown in Fig. 8 (b).
It shows only a bulk hole-like bands without any feature below 0.5 eV, indicating that all the electronic structure near $E_{\rm F}$ observed in this work come from the (2$\times$1)-Bi surface atomic structure.

\end{document}